# Modelling basketball players' performance and interactions between teammates with a regime switching approach*

Paola Zuccolotto[1], Marco Sandri[1], Marica Manisera[1] and Rodolfo Metulini[1]

**Summary**. Basketball players' performance measurement is of critical importance for a broad spectrum of decisions related to training and game strategy. Despite this recognized central role, the main part of the studies on this topic focus on performance level measurement, neglecting other important characteristics, such as variability. In this paper, shooting performance variability is modeled with a Markov Switching dynamic, assuming the existence of two alternating performance regimes. Then, the relationships between each player's variability and the layup composition is modeled as an ARIMA process with covariates and described with network analysis tools, in order to extrapolate positive and negative interactions between teammates, helping the coach to decide the best substitution during the game.

*Keywords*: Sport analytics; Basketball; Performance variability; Markov Switching Model; Network Analysis.

## 1 Introduction

The management of a sport team requires information coming from data and big data. Analytics are essential to support coaches and technical experts in a wide range of decisions. In basketball these decisions are made before the game with reference to training and playing strategies but also during the game, when fast and timely choices have to be made. In this situation, an important subject is to make decisions about which players should be on the court in a given moment. Of course this choice depends on the specific game moment, the team's tactics, the opponent behavior, ... but it is also important to be aware of the mutual interactions of players' performances, in order to exploit possible synergies.

In the scientific literature, players' performance analysis is a hot topic (see for example Page et al. 2007, Cooper et al. 2009, Sampaio et al. 2010, Piette et al. 2010, Fearnhead & Taylor 2011, Ozmen 2012, Page et al. 2013, Erčulj & Štrumbelj 2015, Deshpande & Jensen 2016, Passos et al. 2016, Franks et al. 2016). Some studies deal with a broad concept of performance (accounting for offensive and defensive abilities, for example), others focus just on shot performance, also with reference to the socalled "hot hand" effect (Gilovich et al. 1985, Vergin 2000, Koehler & Conley 2003, Tversky & Gilovich 2005, Arkes 2010, Avugos et al. 2013, Bar-Eli et al. 2006) and with special attention to the impact of high-pressure game situations (Madden et al. 1990, 1995, Goldman & Rao 2012, Zuccolotto et al. 2017). In this paper we address shot performance, but with a special interest on variability.

More specifically, the paper is concerned with the assessment of players' shooting performance - from the twofold perspective of average and variability - and the investigation of its relationships with the team composition and the team performance, with the final aim to give the coach suggestions about the best substitutions during the course of the game. To do that, we propose a three-step procedure:

1. for each player belonging to the same team we describe the cyclical alternation of good and bad performance by means of (*i*) the definition of a proper smoothed index of shooting performance and (*ii*) its analysis with Markov Switching models in order to detect the possible presence of different regimes in performance;

2. for the players with significant evidence of different regimes, we (*i*) measure how the probability of being in a good performance regime is affected by the presence of the other teammates on the court by means of an ARIMA model with covariates (ARIMAX), and (*ii*) we represent all the significant relationships with network analysis graphical tools;

3. we exploit the network obtained in Step 2 to define couples or groups of teammates positively (negatively) influencing one another and check the impact on the team's performance of their joint presence on the filed.

The procedure will be described with reference to a case study using web-scraped freely available play-by-play data from International Basketball Federation (FIBA) web page (www.fiba.basketball.com) concerned with the 20 games played by Iberostar Tenerife team during the 2016/2017 European Champions League.

The paper is organized as follows: Section 2 discuss the literature background that is relevant for our research question, Section 3 introduces the methods and presents the results related to the first step of our process, Section 4 and 5 are devoted to the second step and third step, respectively. Section 6 concludes.

## 2 Background

### 2.1 Performance variability

Human performance evaluation is an important issue in several different contexts, with the prevalent conceptualization being typical performance. Psychological studies have pointed out that typical performance is but one attribute of performance, but other aspects should be taken into account. Sackett et al. (1988) and DuBois et al. (1993) identified the concept of maximal performance, defined as what an individual "can do", to pair with typical performance definition as what an individual "will do". At the same time other research lines have began to focus on variability in performance over time, arguing that performance fluctuation should not to be considered as random noise but is worth of separate analysis (Lecerf et al. 2004) and calling attention to intra-individual differences in variability (Rabbitt et al. 2001). Barnes & Morgeson (2007) applied the above mentioned studies (that were conceived in a general behavioral analysis context) to the environment of sport management. They investigated how the three conceptualizations of performance (typical, maximal and performance variability) affect compensation levels of National Basketball Association (NBA) players, finding strong support to the hypothesis that performance variability would be negatively related to compensation.

Following the arguments put forward by these studies, in this paper we focus on shot performance variability of basketball players. The idea of performance variability should not be confused with the so-called momentum effect, although it is not a completely unrelated issue.

The term psychological momentum is used to describe changes in performance based on success or failure in recent events. Adler (1981) defined psychological momentum - that can be positive or negative according to whether it refers to success or failure - as the tendency of an effect to be followed by a similar effect. Examples of momentum are represented by terms such as hot and cold streaks, the hot hand in basketball shooting, and batting slumps in baseball, that are part of the jargon of sports. Although there have been many attempts to develop models and measures of psychological momentum, and it has been demonstrated that athletes actually perceive that momentum exists, evidence of this effect within individual athletic contexts has proved elusive (Vergin 2000, Bar-Eli et al. 2006, Avugos et al. 2013). Gilovich et al. (1985) studied the hot hand in basketball and explained it as being due to an unfounded belief in a law of small numbers, as later confirmed by other studies (Koehler & Conley 2003, Tversky & Gilovich 2005). Similar cautious conclusions have been reached with reference to several sport, such as baseball (Albright 1993), tennis (Silva III et al. 1988, Richardson et al. 1988), golf (Clark III 2005), although there are also examples of evidences in favor of this phenomenon (see for example Dorsey-Palmateer & Smith 2004 in bowling, Raab et al. 2012 in volleyball and Zuccolotto et al. 2017 in basketball).

In the present study we wish to neither confirm, nor deny the momentum effect. We simply take note that hot and cold streaks do occur, regardless of whether they can be explained by the idea of psychological momentum, and their occurrence generates part of the performance variability over time.

As a matter of fact, different levels of performance may be due to a lot of factors other than psychological ones, such as for example the physical condition and the teammates. In particular, our study of performance variability aims at identifying cyclic patterns of shooting performance and investigating whether they may be associated to the team composition.

## 2.2 Teamwork assessment

In team sports, the overall team performance is not simply given by the sum of the single players' skills, but it is also the result of teamwork. Several studies have considered teamwork assessment from different perspectives. Just to cite a few, Fujimura & Sugihara (2005) construct a player's motion model after proposing a generalized Voronoi diagram that divides space into dominant regions, Metulini et al. (2017*b*), Metulini (2017), Metulini et al. (2017*a*) analysed players' motion using motion charts and proposing a cluster analysis to separate different games' strategies, Vilar et al. (2012) show that interactions between players may be analyzed by ecological dynamics explaining the formation of successful and unsuccessful patterns of play, and Carron & Chelladurai (1981) have attempted to identify the factors correlated with the athlete's perception of cohesiveness (between the coach and athlete and the team and athlete) intended as a multidimensional construct.

Among the methods proposed to deal with teamwork assessment, a noteworthy approach is given by networks methods. Warner et al. (2012) and Lusher et al. (2010) used social network analysis to investigate how team cohesion and individual relationships impact team dynamics, Passos et al. (2011) revealed that the number of network interactions between team members should be able to differentiate between successful and unsuccessful performance outcomes, Clemente et al. (2014) applied a set of network metrics in order to characterize the teammates cooperation in a football team by considering individual analysis, players contribution for the team and global interaction of the team.

In this paper we consider that specific combinations of the players on the court may entail positive or negative synergies and try to highlight these coactions by means of network analysis graphical tools applied to the results of the performance variability analysis.

## 3 Step 1: analysis of shooting performance variability

To measure shooting performance we consider two circumstances: the intensity of shooting and the extent to which shots succeed in scoring baskets. For a given player i, let $\tilde{\phi}_{ij}$ be a measure of his intensity of shooting at shot j,

$$\tilde{\phi}_{ij} = \frac{1}{t_{ij}} \tag{1}$$

where $t_{ij}$ denotes the time elapsed since shot j − 1 of the same player or since the moment when he entered the court. The shooting intensity is partly determined by the opponent's strength, so we remove the match effect and obtain the adjusted shooting intensities

$$\phi_{ij} = \frac{\tilde{\phi}_{ij}}{\bar{\phi}^{(m_{ij})}} \tag{2}$$

where $m_{ij}$ is the match when player i attempted shot j and $\phi^{(m_{ij})} = S^{(m_{ij})}/T$, with $S^{(m_{ij})}$ the total number of shots attempted by the whole team in match $m_{ij}$ and T the duration of the match. Following Zuccolotto et al. (2017), a measure of efficiency of shot j is given by

$$E_{ij} = x_{ij} - p_{ij} \tag{3}$$

where $x_{ij}$ denotes the indicator function assuming value 1 if shot j of player i scored a basket and 0 otherwise, and $p_{ij}$ is the scoring probability of the shot. In this formula, the difference is positive if the shot scored a basket (and the lower the scoring probability, the higher its value) and negative if it missed (and the higher the scoring probability, the higher its absolute value). So, a basket is worth more when the scoring probability of the corresponding shot is low, whereas when a miss occurs, it is considered more detrimental when the scoring probability of the corresponding shot is high (Zuccolotto et al. 2017). To estimate the scoring probability $p_{ij}$ we use the goal percentage statistics of match $m_{ij}$ (so, we remove the possible match effect), separately for 2-points (2P) and 3-points (3P) shots, and free throws (FT).

The measures $\phi_{ij}$ and $E_{ij}$ are event-level, *i.e.* they are computed for each shot. On this point, it's worth considering some remarks. In general, the studies of intraindividual variability tend to examine performance over preaggregated periods, comparing weeks, semesters, or years. As noted by Diener & Larsen (1984), aggregating data in this fashion leads to more stable and consistent estimates than those based on disaggregated data, but at the same time this may mask episodic variation in performance. On the other hand, investigating performance variability at the event-level will avoid the masking effects just mentioned (Barnes & Morgeson 2007), but event-level measures are often characterized by a high level of noise that may hidden structural relationships. For these reasons, we opt for the middle ground solution of averaging the measures over short moving periods, by means of the Nadaraya-Watson kernel regression (Nadaraya 1964, Watson 1964), estimated with a narrow bandwidth. More specifically, we use a Gaussian kernel and set the bandwidth, for each player, equal to the 25th percentile of his number of shots per match, in order to obtain intra-match averages. We denote with $\hat{\phi}_{ij}$ the smoothed estimates of the shooting intensities. A unique measure of shooting performance $\psi_{ij}$ is then obtained as

$$\psi_{ij} = \hat{\phi}_{ij} E_{ij}. \tag{4}$$

The rationale behind formula (4) is that the shooting intensities can amplify or shrink the shot efficiencies: a moment when the player's shots tend to be highly efficient results in a higher performance if also the shooting intensity is high in the same moment, and vice versa. Following the reasoning described above, also the measures $\psi_{ij}$ have been smoothed by means of the same the Nadaraya-Watson kernel regression and we denote with $\hat{\psi}_{ij}$ the corresponding smoothed estimates. Figures 1 and 2 show examples of the describes measures for one selected player of the Team "Iberostar Tenerife", whose performance analysis in the Basketball Champions League 2016/17 we report as a case study, as specified in the Introduction. The vertical dashed lines divide consecutive matches. One match has been deleted from the dataset due to missing data about the layup composition.

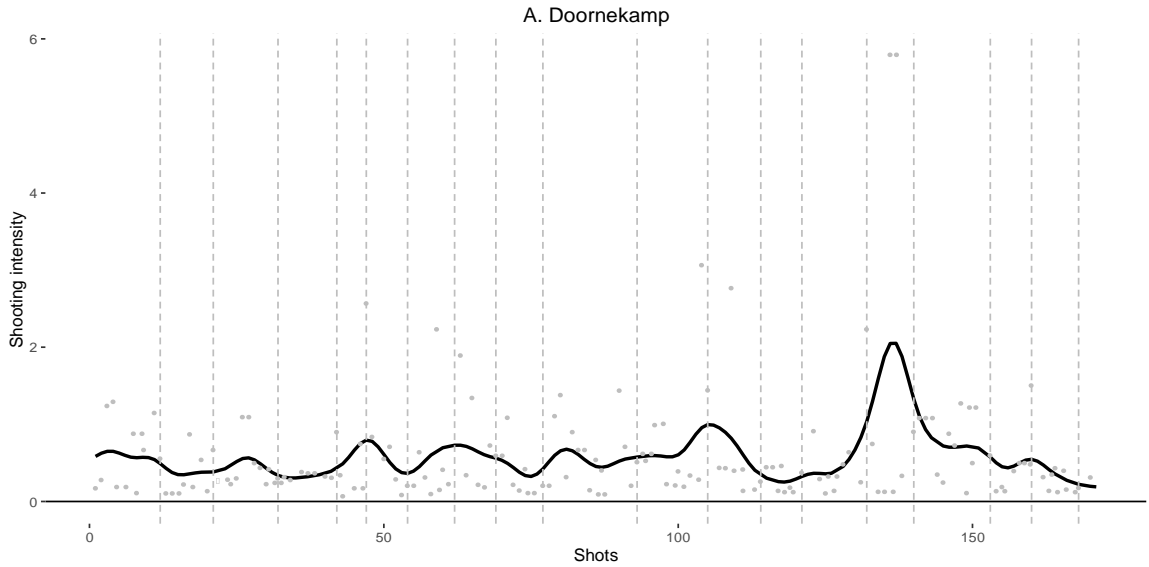

Figure 1: Shooting Intensity: measures $\phi_{ij}$ (gray points) and $\hat{\phi}_{ij}$ (black line)

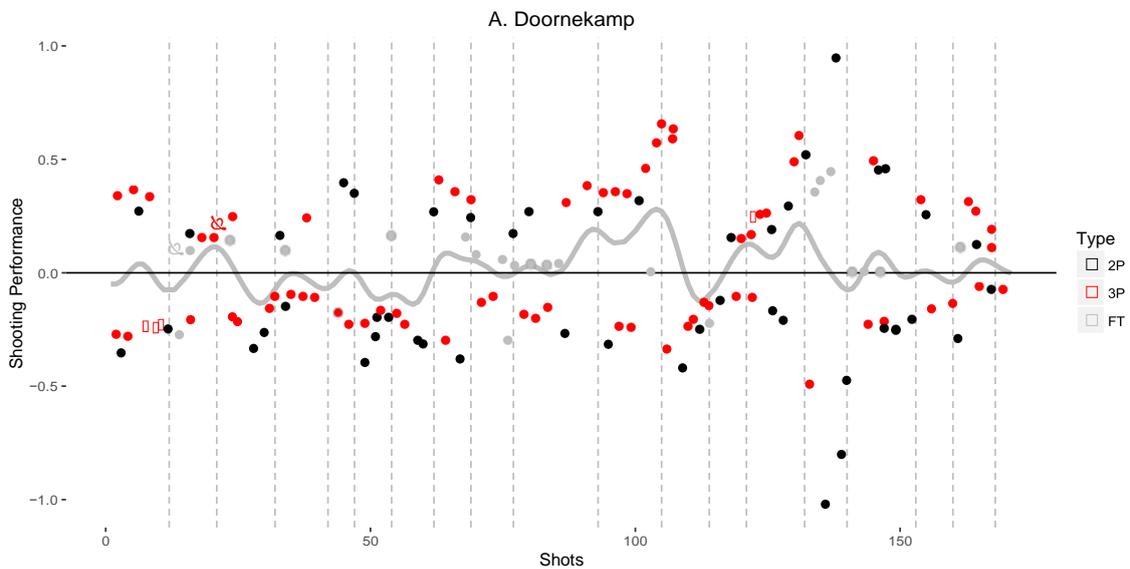

Figure 2: Shooting Performance: measures $\psi_{ij}$ (coloured points, with different colour according to the type of shot) and $\hat{\psi}_{ij}$ (gray line)

For each player we consider the mean and the standard deviation of the shooting performance, $\Psi_i = E(\hat{\psi}_{ij})$ and $\sigma_i = \sqrt{\overline{E[(\hat{\psi}_{ij} - \Psi_i)^2]}}$, estimated respectively by $\text{av}_j(\hat{\psi}_{ij})$ and $\text{sd}_j(\hat{\psi}_{ij})$, where $\text{av}_j(\cdot)$ and $\text{sd}_j(\cdot)$ denote the average and the standard deviation over j. In Figure 3 the players are plotted as bubbles with coordinates $(\text{av}_j(\hat{\psi}_{ij}), \text{sd}_j(\hat{\psi}_{ij}))$ and size equal to the total number of shots.

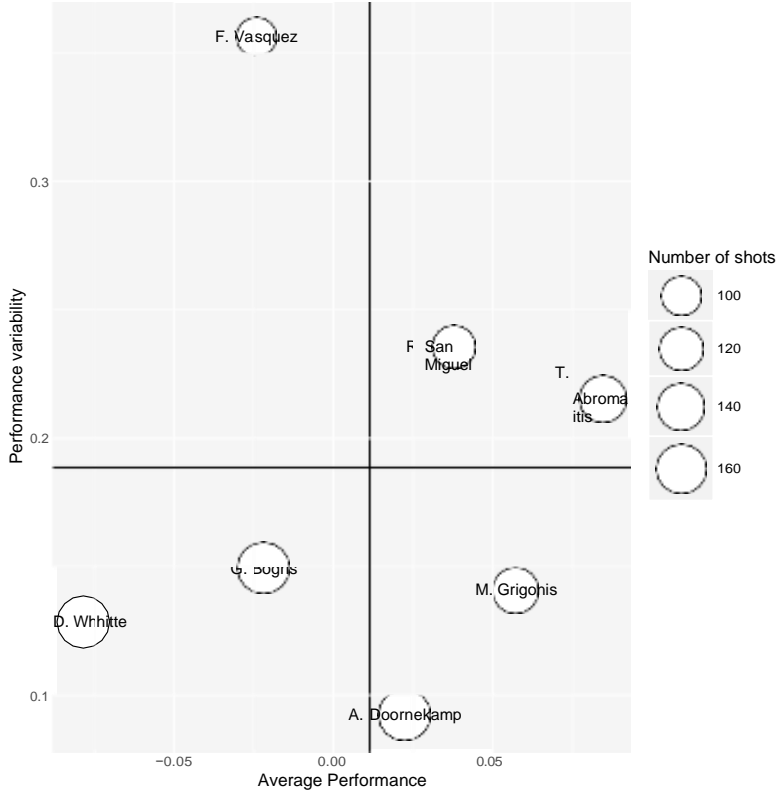

Figure 3: Average performance vs. Performance variability

## 3.1 Markov switching modelling of two-regimes performance

We now consider the smoothed performance measures $\hat{\psi}_{ij}$ and their variability. Specifically, we aim to determine whether their fluctuating pattern may be considered simply erratic, or we may instead recognize an alternation of good and bad performance over time. To do that we resort to Markov Switching models (see Hamilton 2010), mainly used in time series econometrics, but fairly adaptable to this context, except the only difference that here we deal with event-time other than clock-time. Let us assume that the performance of player i switches between two different regimes, G and B ("good" and "bad", respectively), where G means that his performance is improving, and B means that it is worsening. In order to argue in terms of variations (improvement or worsening), we consider the values or $\hat{\psi}_{ij,\Delta}(k) = \text{sgn}(\hat{\psi}_{ij} - \hat{\psi}_{i(j-1)}) \cdot |\hat{\psi}_{ij} - \hat{\psi}_{i(j-1)}|^k$ where k is a correction factor used to amplify or shrink differences. If $k = 1$, we have $\hat{\psi}_{ij,\Delta}(1) = \hat{\psi}_{ij} - \hat{\psi}_{i(j-1)}$, *i.e.* first order differences. It's worth noting that since the values $|\hat{\psi}_{ij} - \hat{\psi}_{i(j-1)}|$ are usually lower than 1, differences are emphasized by using $k < 1$. Let $R_{ij}$ be the (unobserved) random variable denoting the regime of player i's performance when he attempted shot j:

$$E(\hat{\psi}_{ij,\Delta}(k)|R_{ij} = r) = \Psi^r_i \qquad r = G, B. \tag{5}$$

The probabilistic model describing the regimes dynamics is assumed to be a two-state Markov chain

(Baum et al. 1970, Lindgren 1978, Hamilton 1989),

$$\Pr(R_{ij}|R_{i(j-1)}, R_{i(j-2)}, \ldots) = \Pr(R_{ij}|R_{i(j-1)}) \tag{6}$$

and we denote with $\pi_{iGG} = \Pr(R_{ij} = G|R_{i(j-1)} = G)$ and $\pi_{iBB} = \Pr(R_{ij} = B|R_{i(j-1)} = B)$ the two-state transition probabilities for player i, recalling that $\pi_{iBG} = \Pr(R_{ij} = B|R_{i(j-1)} = G) = 1 - \pi_{iGG}$ and $\pi_{iGB} = \Pr(R_{ij} = G|R_{i(j-1)} = B) = 1 - \pi_{iBB}$.

After specifying Gaussian densities N ($\Psi_i^G$, $\sigma_i(G)^2$) and N ($\Psi_i^B$, $\sigma_i(B)^2$) under the two regimes, the parameter vector $\theta_i = (\Psi_i^G, \Psi_i^B, \sigma_i(G)^2, \sigma_i(B)^2, \pi_{iGG}, \pi_{iBB})'$ is estimated via EM algorithm (we used the R package MSwM, Sanchez-Espigares & Lopez-Moreno 2014), as the regime is unobserved. The estimation algorithm also returns the socalled "filltered" probabilities

$$\pi_{ijr|j} = \Pr(R_{ij} = r|I_j, \theta_i) \tag{7}$$

where $I_j$ denotes the information available up to shot j, and the corresponding "smoothed" probabilities

$$\pi_{ijr} = \Pr(R_{ij} = r|I, \theta_i) \tag{8}$$

obtained using all the set of information I up to the last shot, by means of the algorithm developed by Kim (1994).

We fitted the described model to all the players that attempted at least 100 shots. We used different values of $k \in (0, 1.5]$ and, for each k, we recorded which players exhibit significant regime switching dynamic (*i.e.* we have $\Psi_i^G > 0$ and $\Psi_i^B < 0$ and both parameter are significant at the 95% confidence level). We found that with $k \leq 0.6$ all the players' dynamics are considered regime switching. Figure 4 shows which players' dynamics continue to be significantly switching as k increases and gives us suggestion for a sort of ranking of players from this point of view. In particular, the player most subject to regime switching turns out to be Doornerkamp (significantly switching with $k \leq 1.3$), followed by Grigonis (k ≤ 1), White (k ≤ 0.9), Vazquez (k ≤ 0.8) and all the others (k ≤ 0.6).

In order to consider in the analysis all the players, we fixed k = 0.5. The corresponding parameter estimates are displayed in Table 1, while Figure 5 shows an example of the estimated regime switching.

If the Markov chain is presumed to be ergodic, we can compute the unconditional probabilities $\pi_{ir}$ (r = G, B) of the two regimes,

$$\pi_{iG} = \frac{1 - \pi_{iBB}}{2 - \pi_{iGG} - \pi_{iBB}} \quad \text{and} \quad \pi_{iB} = 1 - \pi_{iG} \tag{9}$$

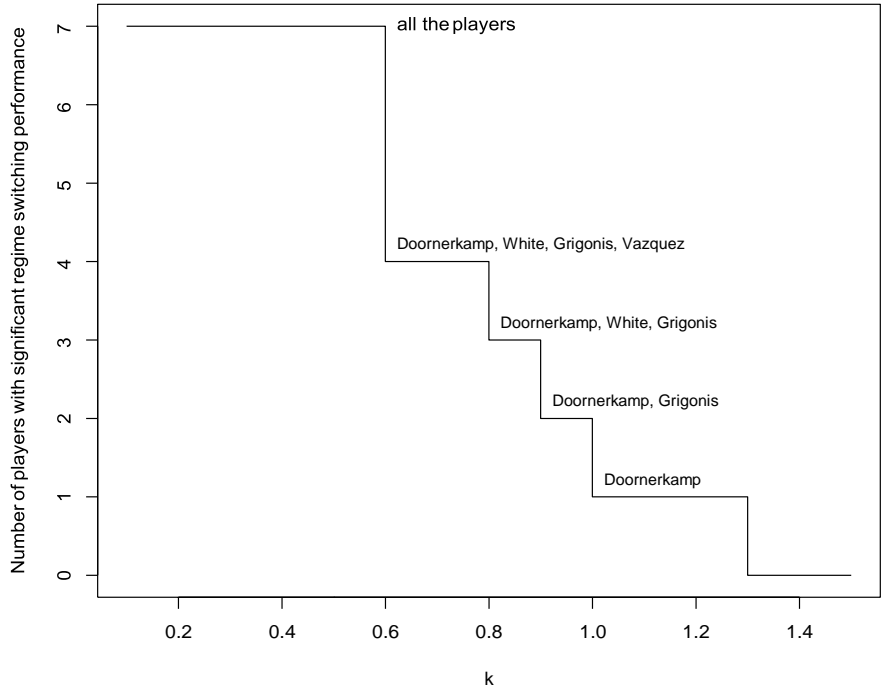

Figure 4: Players with significant regime switching for different values of k

Table 1: Parameter estimates of Markov Switching models

| Player | $\Psi_i^G$ | $\Psi_i^B$ | $\sigma_i(G)^2$ | $\sigma_i(B)^2$ | $\pi_{iGG}$ | $\pi_{iBB}$ |
|---|---|---|---|---|---|---|
| A. Doornekamp | 0.1410 | -0.1274 | 0.0026 | 0.0039 | 0.8288 | 0.8429 |
| D. White | 0.1368 | -0.1303 | 0.0036 | 0.0040 | 0.9137 | 0.9187 |
| G. Bogris | 0.1498 | -0.1297 | 0.0054 | 0.0055 | 0.8410 | 0.8271 |
| M. Grigonis | 0.1593 | -0.1521 | 0.0045 | 0.0053 | 0.8624 | 0.8702 |
| R. San Miguel | 0.1912 | -0.2293 | 0.0302 | 0.0251 | 0.8243 | 0.7806 |
| T. Abromaitis | 0.218 | -0.1517 | 0.0116 | 0.0288 | 0.8151 | 0.8703 |
| F. Vazquez | 0.2671 | -0.3373 | 0.0566 | 0.0268 | 0.7621 | 0.6699 |

and their average persistence $\delta_{ir}$ (*i.e* the average duration, expressed as number of shots)

$$\delta_{iG} = \frac{1}{1-\pi_{iGG}} \quad \text{and} \quad \delta_{iB} = \frac{1}{1-\pi_{iBB}}. \tag{10}$$

Table 2 display the values obtained for the selected players.

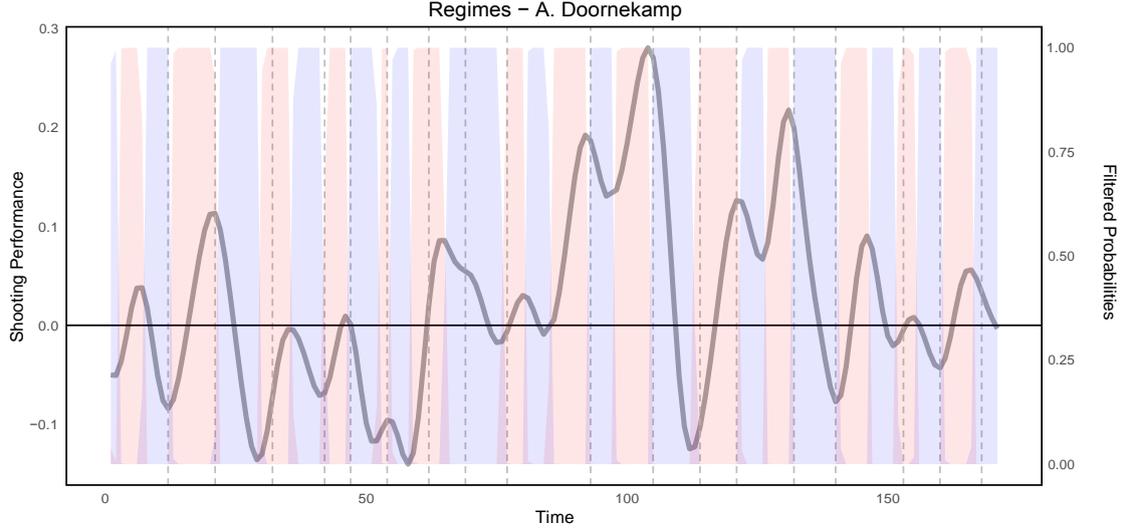

Figure 5: Regime Switching: measures $\hat{\psi}_{ij}$ (gray line) with filtered probabilities $\pi_{ijG|j}$ (red area) and $\pi_{ijB|j}$ (blue area)

Table 2: Unconditional probability of the regime with good performance and average persistence of the regimes

| Player | $\pi_{iG}$ | $\delta_{iG}$ | $\delta_{iB}$ |
|---|---|---|---|
| A. Doornekamp | 0.4785 | 5.8411 | 6.3654 |
| D. White | 0.4851 | 11.5875 | 12.3001 |
| G. Bogris | 0.5209 | 6.2893 | 5.7837 |
| M. Grigonis | 0.4854 | 7.2674 | 7.7042 |
| R. San Miguel | 0.5553 | 5.6915 | 4.5579 |
| T. Abromaitis | 0.4123 | 5.4083 | 7.7101 |
| F. Vazquez | 0.5812 | 4.2034 | 3.0294 |

# 4 Step 2: assessment of team interactions and visualization with graphical network analysis

In the second step, we consider the interactions among players by assessing the extent to which the presence on the court of a teammate player h affects the probability of player i of being in the regime with good performance. Let $C_{ijh}$ be a dichotomous random variable assuming value 1 if player h is on the court when player i attempts shot j, and 0 otherwise. The regimes of player i are independent on the presence of teammate h if

$$\Pr(R_{ij} = r | C_{ijh} = 0) = \Pr(R_{ij} = r | C_{ijh} = 1) = \pi_{ir}. \quad (11)$$

The regime probabilities $\Pr(R_{ij} = r)$ are estimated by means of the filtered or smoothed probabilities defined respectively in (7) and (8), but to measure their association to $C_{ijh}$ we have to take into account that a strong autocorrelation structure is present. For each couple of players i and h, we estimated the

ARIMAX model

$$\pi_{ijG|j} = \varepsilon_j + \sum_{l=1}^{p} \alpha_l \pi_{i(j-l)G|j-l} + \sum_{l=1}^{q} \gamma_l \varepsilon_{j-l} + \beta_{ih} C_{ijh} \tag{12}$$

where d-order differencing is applied when necessary. The orders p, d and q of the models have been identified by means of the automatic stepwise procedure based on information criteria described in Hyndman et al. (2007) and implemented in the R package forecast. Table 3 displays the parameters $\beta_{ih}$ with p-values lower than 0.1 and 0.05.

Table 3: Significant (90%) parameters $\beta_{ih}$ (rows: players i - columns: teammates h). * = p-value < 0.05

|  | A. Doornekamp | D. White | G. Bogris | M. Grigonis | R. San Miguel | T. Abromaitis | F. Vazquez |
|---|---|---|---|---|---|---|---|
| A. Doornekamp | - | 0.0880* | - | -0.0623 | -0.0844 | 0.0673 | 0.0735 |
| D. White | - | - | - | - | - | - | - |
| G. Bogris | 0.0856 | - | - | - | - | - | - |
| M. Grigonis | - | - | - | - | - | - | - |
| R. San Miguel | - | -0.0667 | - | - | - | - | 0.1013* |
| T. Abromaitis | - | - | - | -0.0801 | - | - | - |
| F. Vazquez | 0.1202* | - | - | - | 0.1406* | - | - |

Figure 6 gives a graphical representation of the interactions depicted by the parameters $\beta_{ih}$. The nodes of the networks are the players and the node size is proportional to the total number of shots attempted by each player. The (directed) edges represent the influence of teammate h on the shooting performance regime of player i, blue and red edges denote respectively negative and positive influence ($\beta_{ih} < 0$ and $\beta_{ih} \geq 0$), and their thickness is proportional to $|\beta_{ih}|$. The left panel shows the network obtained using the parameters significant at the 90% level, while the right panel restricts to parameters significant at the 95% level.

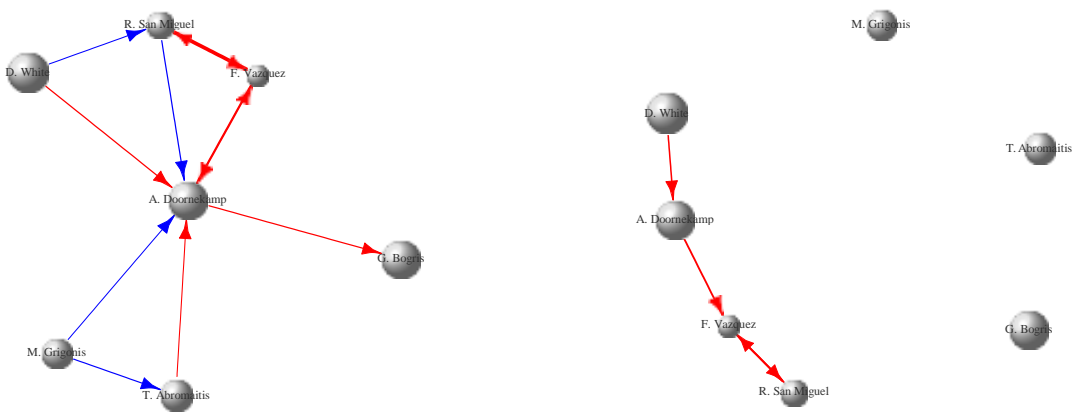

Figure 6: Influence of teammates on shooting performance (left: 90% significance level; right: 95% significance level)

Some network measures can be computed. To this aim we use igraph package in R and refer to the graph on the left panel. The density (Wasserman & Faust 1994) of this network is 0.2619 (the density is 1 when every player is connected with everybody else). Since the graph is directed, active edges could go from player A to player B, from player B to player A or in both directions. Reciprocity index measures to what extent the links are reciprocal (i.e. if the presence on the court of player A influences players' B shooting performance, the relation holds in the other way round as well) or not. The index stands to 0.3636. We then computed eigenvector centrality, scaled to 1 (Bonacich 1987), for the seven analysed players. Doornekamp and Vazquez are the two most central players, with a scaled centrality measure of 1. Bogris and San Miguel follow, with the same measure (0.6180), while White, Grigonis and Abromaitis report a centrality of 0. Then, to better characterize the centrality structure, we compute total- , in- and out-degree coefficients for each player. Doornekamp has the larger total-degree coefficients (first column of Table 4), San Miguel and Vasquez follow with 4. The total-degree coefficient measures how many edges start from (or go to) the specific vertex and can give a hint about which player is more important than others in the network. The total-degree does not distinguish whether the player influences the others or is influenced by them, and this information is carried by in- and out-degree coefficients (columns 2 and 3). Doornekamp exhibits the larger in-degree coefficient, meaning that his shooting performance is (positively or negatively) the mostly affected by teammates, followed by San Miguel and Vazquez. In terms of out-degree, five players present the maximum value (2).

Table 4: Degree measures for the seven analysed players (in brackets, the number of positive and negative edges).

| player | Degree | | | Strength degree | |
|---|---|---|---|---|---|
| | total | in (+/-) | out (+/-) | in | out |
| Doornekamp | 7 | 5 (3/2) | 2 (2/0) | 0.0821 | 0.2058 |
| White | 2 | 0 (0/0) | 2 (1/1) | 0.0000 | 0.0213 |
| Grigonis | 2 | 0 (0/0) | 2 (0/2) | 0.0000 | -0.1424 |
| San Miguel | 4 | 2 (1/1) | 2 (1/1) | 0.0346 | 0.0562 |
| Abromaitis | 2 | 1 (0/1) | 1 (1/0) | -0.0801 | 0.0673 |
| Vazquez | 4 | 2 (2/0) | 2 (2/0) | 0.2608 | 0.1748 |
| Bogris | 1 | 1 (1/0) | 0 (0/0) | 0.0856 | 0.0000 |

In order to take into account the signs of the edges we computed the in- and out-strength degree coefficients (Barrat et al. 2004, columns 4 and 5), weighting the simple degree measures by the significant coefficients of the ARIMAX model (Table 3) to adjust for the presence of positive/negative effects. Doornekamp and Vazquez exhibit positive (in- and out-) strength degree coefficients, while, for example, Grigonis has a negative out-strength degree (his presence on the court negatively affects Doornekamp and Abromaitis shooting performance). All in all, it looks like Doornekamp and Vazquez are the players that mostly affect other players' shot performance, as confirmed by different network measures.

# 5 Step 3: impact on the team performance

The aim of Step 3 is to check the relationships between teammates interactions and team performance. In other words, we answer the following question: does the presence of specific player, or the jointly presence on the court of a group of players - identified relying on the network analysis of Step 2 - (positively or negatively) affect team performance, intended as the points scored by the team?

To answer this question, we exploit the information coming from the network analysis of Step 2 to select subsamples of the games and compare them in terms of a measure related to the intensity of scored points. The selection of subsamples is made relying on three different criteria:

1. The presence on the court of Doornekamp, the player exhibiting the highest values of the network indexes; we compare the intensity of scored points of this subsample to that of all the other moments.

2. The jointly presence on the court of couples of players, chosen in such a way that the first one positively affects the second (in the sense defined in Step 2): White/Doornekamp, Doornekamp/Vazquez, Vazquez/San Miguel, San Miguel/Vazquez (right panel of Figure 6); we compare the intensity of scored points of each subsample (*i.e.* when both players of a given couple are on the court) to the moments when only the second player of the couple is on the court.

3. a more complex rule considering both positive an negative interactions among all the players (left panel of Figure 6), as will be better specified later (Table 7).

The intensity of scored point, ISP, is defined as follows.

Let W be the random variable denoting the points scored in a given game second and $w_t$ ($w_t = 0, 1, 2, 3, 4$) the value assumed by W at time (second) $t$ ($t = 1, 2, \ldots, 2400$). Note that the points scored with free throws are assigned to the time of the corresponding foul.

Let $T$ be the total time of the game, composed of all the $s(T) = 2400$ game seconds and $\tau \subseteq T$ a subsample, defined according to a specified criterion, containing $s(\tau) \leq s(T)$ seconds.

We define ISP in the subsample $\tau$ as

$$\text{ISP}(\tau) = \frac{1}{s(\tau)} \sum_t w_t \qquad (13)$$

ISP($\tau$) can be transformed is order to be interpreted with reference to a standard period of 40 minutes, into the corresponding index $\text{ISP}_{40}(\tau) = \text{ISP}(\tau) \times 2400$.

Results are summarized is Tables 5 - 7 for the three criteria, respectively.

The differences $\text{ISP}_{40}(\tau_1) - \text{ISP}_{40}(\tau_2)$ are all positive, meaning that the mutual impacts of teammates designed by the network analysis applied to the Markov Switching performance regimes have an effective association with the points scored by the team.

Table 5: Comparison of $ISP_{40}$ indexes - Criterion 1.

| Selected subsample ($\tau_1$) | Comparison subsample ($\tau_2$) | $ISP_{40}(\tau_1) - ISP_{40}(\tau_2)$ |
|---|---|---|
| when Doornekamp is on the court | when Doornekamp is not on the court ($\tau_1 \cup \tau_2 = T$, $\tau_1 \cap \tau_2 = \emptyset$) | 5.65 |

Table 6: Comparison of $ISP_{40}$ indexes - Criterion 2.

| Selected subsample ($\tau_1$) | Comparison subsample ($\tau_2$) | $ISP_{40}(\tau_1) - ISP_{40}(\tau_2)$ |
|---|---|---|
| when White and Doornekamp are both on the court | when only Doornekamp is on the court ($\tau_1 \cup \tau_2 \subseteq T$, $\tau_1 \cap \tau_2 = \emptyset$) | 3.33 |
| when Doornekamp and Vazquez are both on the court | when only Vazquez is on the court ($\tau_1 \cup \tau_2 \subseteq T$, $\tau_1 \cap \tau_2 = \emptyset$) | 12.86 |
| when Vazquez and San Miguel are both on the court | when only San Miguel is on the court ($\tau_1 \cup \tau_2 \subseteq T$, $\tau_1 \cap \tau_2 = \emptyset$) | 19.10 |
| | when only Vazquez is on the court ($\tau_1 \cup \tau_2 \subseteq T$, $\tau_1 \cap \tau_2 = \emptyset$) | 9.54 |

Table 7: Comparison of $ISP_{40}$ indexes - Criterion 3.

| Selected subsample ($\tau_1$) | Comparison subsample ($\tau_2$) | $ISP_{40}(\tau_1) - ISP_{40}(\tau_2)$ |
|---|---|---|
| Doornekamp and Vazquez are both on the court, but San Miguel (who negatively affects Doornekamp) is not on the court | Doornekamp, Vazquez and San Miguel are all on the court ($\tau_1 \cup \tau_2 \subseteq T$, $\tau_1 \cap \tau_2 = \emptyset$) | 1.34 |
| Abromaitis and Doornekamp are both on the court, but Grigonis (who negatively affects both of them) is not on the court | Abromaitis, Doornekamp and Grigonis are all on the court ($\tau_1 \cup \tau_2 \subseteq T$, $\tau_1 \cap \tau_2 = \emptyset$) | 0.46 |
| White, Doornekamp, Vazquez and San Miguel are all on the court | at least one of them is not on the court ($\tau_1 \cup \tau_2 \subseteq T$, $\tau_1 \cap \tau_2 = \emptyset$) | 8.90 |

# 6 Concluding remarks

In this paper we have proposed a three-step procedure aimed at giving to basketball coaches a tool supporting their decisions about which players should be jointly on the court in a given moment. A real data case study developed using play-by-play data from the 2016/2017 European Champions League has been analyzed alongside the methodological definition of the procedure. Firstly, starting from the idea that players' performance is naturally subject to rise and fall, we have defined a shooting performance index and we have modeled its alternating dynamic with Markov Switching models, borrowed from econometrics. Secondly, we have modeled the probability of being in a "good performance regime" (rise) with respect to the presence of teammates on the court, finding out positive and negative interactions between players from this point of view. To do that, we resorted to an ARIMA model with covariates and networks analysis tools to graphically describe the emerged relationships. Finally, we checked whether these relationships among players effectively translate into a better team performance, from the point of view of the scored points and found confirmation in all the examined cases.

**Acknowledgments**

Research carried out in collaboration with the Big&Open Data Innovation Laboratory (BODaI-Lab), University of Brescia (project nr. 03-2016, title "Big Data Analytics in sports").